\documentclass[aps,preprint]{revtex4}%
\usepackage{amsfonts}
\usepackage{amsmath}
\usepackage{amssymb}
\usepackage{graphicx}%
\begin{document}
\title{Correlation, Entropy, and Information Transfer in Black Hole Radiation}
\author{Baocheng Zhang$^{a}$}
\email{zhangbc@wipm.ac.cn}
\author{Qing-yu Cai$^{a}$}
\email{qycai@wipm.ac.cn}
\author{Ming-sheng Zhan$^{a}$}
\email{mszhan@wipm.ac.cn}
\author{Li You$^{b}$}
\email{lyou@mail.tsinghua.edu.cn}
\affiliation{$^{a}$State Key Laboratory of Magnetic Resonances and Atomic and Molecular
Physics, Wuhan Institute of Physics and Mathematics, The Chinese Academy of
Sciences, Wuhan 430071, China}
\affiliation{$^{b}$Department of Physics, State Key Laboratory of Low Dimensional Quantum
Physics, Tsinghua University, Beijing 100084, China}
\begin{abstract}
Since the discovery of Hawking radiation, its consistency with quantum theory
has been widely questioned. In the widely described picture, irrespective of
what initial state a black hole starts with before collapsing, it eventually
evolves into a thermal state of Hawking radiations after the black hole is
exhausted. This scenario violates the principle of unitarity as required for
quantum mechanics and leads to the acclaimed \textquotedblleft information
loss paradox\textquotedblright. This paradox has become an obstacle or a
reversed touchstone for any possible theory to unify the gravity and quantum
mechanics. Based on the results from Hawking radiation as tunneling, we
recently show that Hawking radiations can carry off all information about the
collapsed matter in a black hole. After discovering the existence of
information-carrying correlation, we show in great detail that entropy is
conserved for Hawking radiation based on standard probability theory and
statistics. We claim that information previously considered lost remains
hidden inside Hawking radiation. More specifically, it is encoded into
correlations between Hawking radiations. Our study thus establishes harmony
between Harking radiation and the unitarity of quantum mechanics, which
establishes the basis for a significant milestone towards resolving the
long-standing information loss paradox. The paper provides a brief review of
the exciting development on Hawking raidation. In addition to summarize our
own work on this subject, we compare and address other related studies.

\end{abstract}
\maketitle


\section{Introduction}

Since the discovery of thermal radiation emissions from a black hole by
Hawking \cite{swh74,swh75}, the foundation of modern physics experiences a
serious challenge as the thermal spectrum is inconsistent with the unitary
principle of quantum mechanics \cite{swh76}. A thermal spectrum implies
information about matter collapsed into a black hole is lost in the process of
black hole radiation. Many have attempted
\cite{acn87,kw89,jp92,jdb93,ck01,hm04,sw05,bp07} to find a resolution but
failed. This lack of a complete resolution makes the paradox of Hawking
radiation an attractive topic up until now. In fact, each failed attempt for a
resolution seems to have made the existence of this paradox more serious and
attracted more interest. In particular, after the possibility that information
about infallen matter may hide inside correlations between Hawking radiation
and the internal states of a black hole was ruled out \cite{bp07}, it seemed
either unitarity or Hawking's semiclassical treatment of radiation must break down.

An important reason for information loss in black hole radiation is usually
attributed to the thermal spectrum, which was obtained by Hawking in his
original calculation assuming a fixed background geometry without enforcing
energy conservation \cite{swh74,swh75,swh76}. A thermal spectrum rules out the
existence of correlations among Hawking radiations, consequently the entropy
for the whole system consisting of the black hole and its radiations must
increase \cite{whz82,dnp82}. However, the entropy for a closed system governed
by quantum mechanics cannot increase. The thermal radiation from a black hole
discovered by Hawking thus causes a serious conflict between general
relativity and quantum mechanics.

A glimpse of hope was raised by the work \cite{jdb93} of Bekenstein, who noted
that information leakage from a black hole is possible if the emitted
radiations contain a subtle non-thermal correction instead of being pure
thermal. This enlightened the path for a resolution to the paradox. A series
of recent studies \cite{sdm09,sdm11,sdm112} claim that the paradox between
black hole radiation and unitarity of quantum mechanics seem inevitable, even
with the small correction to Hawking's leading order calculation. These
discussions, however, can only support a judgement that a possible reason for
the breakdown of unitarity based on the result of a semiclassical calculation
of the black hole radiation lies at the assumption of a fixed spacetime
background. Although analysis about whether a small non-thermal correction
could save the lost information were made in these papers
\cite{sdm09,sdm11,sdm112}, their assumption of a fixed spacetime background
made the physical mechanism from a small correction unclear and consequently
the small corrections discussed in these studies are trivial.

Recently, Parikh and Wilczek \cite{pk00} treated Hawking radiation as
tunneling while enforcing energy conservation. Their method goes beyond a
fixed spacetime background and include back reaction from the emission. As a
result the emission spectrum obtained becomes non-thermal. Starting with this
non-thermal spectrum, we recently visited and revisited the paradox of black
hole information loss \cite{zcyz09,zczy11,zczy112}. Correlations are
discovered to exist among Hawking radiations for all types of black holes
\cite{zczy11,zczy112}. Upon carefully evaluating the amount of information
carried away in these correlations, we find that information is not lost and
the entropy for the total system of a black hole plus its radiations is
conserved, i.e., information coded into the correlations and carried away by
the Hawking radiation is found to match exactly the amount previously claimed
lost. Our conclusion thus restores consistency of Hawking radiation with the
principle of unitarity for quantum mechanics concerning an isolated system.

In this brief review, we will introduce and summarize our recent results in a
more logical and explicit approach. Several important elements required to
understand the information loss paradox and our proposed resolution will also
be discussed. The rest of this paper is organized as follows. We will discuss
the reaction and the non-thermal radiation spectrum obtained using the method
of Hawking radiation as tunneling in the second section. This is followed in
the third section by the description of our earlier discovery of correlations
among Hawking radiations and entropy conservation in the process of Hawking
radiation as tunneling. With the total entropy conserved, a question naturally
appears: why does a black hole has entropy? In the fourth section, we try to
explain this by presenting the meaning of entropy for a black hole from the
micro perspective of Hawking radiation. The fifth section concerns the
transfer of information hidden in a black hole. Finally, we summarize and
conclude in the sixth section.

\section{Reaction and Hawking Radiation Spectrum}

The back reaction of emission is first considered in the calculation of black
hole radiation as tunneling by Parikh and Wilczek \cite{pk00}. The
achievements and prospects for the tunneling method and its applications to
Hawking radiation are documented in a recent topical review \cite{vac11},
which also include discussions on some of our recent work.

In the tunneling method, the Schwarzschild coordinates
\begin{equation}
ds^{2}=-\left(  1-\frac{2M}{r}\right)  dt^{2}+\left(  1-\frac{2M}{r}\right)
^{-1}dr^{2}+r^{2}d\Omega^{2}\label{sm}%
\end{equation}
are usually transformed into the Painlev\'{e} coordinates
\begin{equation}
ds^{2}=-\left(  1-\frac{2M}{r}\right)  dt^{2}+2\sqrt{\frac{2M}{r}}%
dtdr+dr^{2}+r^{2}d\Omega^{2},\label{pm}%
\end{equation}
which are stationary and non-singular at the horizon $r=2M$ and present a
background to define an effective vacuum state for a quantum field essentially
for a free falling observer through the horizon. A second advantage for the
Painlev\'{e} coordinates arises in the Hamilton-Jacobi method by the
calculation of the temperature without the requirement of introducing a
normalization \cite{zcz09}. In the Parikh-Wilczek tunneling method, the total
mass is fixed while the mass of the black hole is allowed to fluctuate, and
the particle of energy $\mathit{E}$ travels on its geodesics which is easily
from the metric (\ref{pm}). Considering the self-gravitation effect, the
metric becomes%
\begin{equation}
ds^{2}=-\left(  1-\frac{2\left(  M-\mathit{E}\right)  }{r}\right)
dt^{2}+2\sqrt{\frac{2\left(  M-\mathit{E}\right)  }{r}}dtdr+dr^{2}%
+r^{2}d\Omega^{2},\label{pm1}%
\end{equation}
and the outgoing radial null geodesics of an emitted particle with energy
$\mathit{E}$ is given by%
\begin{equation}
\overset{.}{r}=1-\sqrt{\frac{2\left(  M-\mathit{E}\right)  }{r}},
\end{equation}
where the ingoing radial null geodesics is obtained by replacing $1$ with $-1$
with the implicit assumption that $t$ increases towards the future.

Due to the infinite blueshift near the event horizon, the WKB approximation
can be employed to calculate the tunneling probability. The calculation
details can be found in the paper \cite{pk00} and will not be reproduced here.
We will instead only give the result for the tunneling probability
\begin{equation}
\Gamma\left(  \mathit{E}\right)  \sim\exp\left[  -8\pi\emph{E}\left(
M-\frac{\mathit{E}}{2}\right)  \right]  =\exp\left(  \Delta S\right)  ,
\label{nt}%
\end{equation}
where $S=4\pi M^{2}$ is the black hole's Bekenstein-Hawking entropy. This
method of calculation enforces energy conservation, or the reaction from
Hawking radiation. The emission spectrum is found to be non-thermal, dependent
on the energy of tunneling particle. Although the tunneling mechanism is used,
the black hole radiation is still understood as stemming from quantum
fluctuations of virtual particles near the event horizon. After a pair of
virtual particles is created at the outside of the black hole, the particle
with negative energy tunnels into the black hole with probability $\Gamma$,
while the particle with positive energy flies away and will be observed at
infinity as black hole radiation. Since particles with negative energy might
not tunnel into the black hole completely, the remainder are annihilated by
their counterparts (particles with positive energy) at the outside of the
black hole. This assures particles with positive energy fly away with the same
tunneling probability $\Gamma$, which is also the probability for black hole
radiation. Sometimes, tunneling are explained inside and near the event
horizon of a black hole with a similar correspondence, but whether quantum
fluctuations of the interior of a black hole is the same as that described by
the common quantum field theory is unclear.

We note that in one of the final remarks of the paper \cite{pk00}, Parikh and
Wilczek speculate about the possibility of information-carrying correlations
in the non-thermal radiation spectrum (\ref{nt}). A recent paper \cite{sdm09},
however, presents a theorem that states the opposite. It claims that for a
traditional black hole or the Schwarzschild black hole, information about
collapsed matter would be inevitably lost in the semiclassical consideration
of Hawking radiation, even including the small corrections to the leading
order calculation by Hawking. Otherwise other paradoxes like infinite
entanglement caused by remnants would appear. While this theorem is
self-consistent within its own frame of a fixed spacetime background, it is
different from the considerations leading to the non-thermal spectrum which
includes the reaction to the spacetime \cite{czzy12}. It is interesting to
note that information loss paradox by a fixed spacetime background has caused
a speculation for gauge-gravity duality \cite{mp13} as a powerful tool to
explore quantum gravity, even in anti-de Sitter space.

As alluded to in the above, we introduce in the following, Hawking radiation
as tunneling invokes pair creation before tunneling and modifications to a
fixed spacetime background using the back reaction from emission. The emission
spectrum consequently becomes non-thermal. For the increased dimensionality of
the interior Hilbert space of a black hole due to the in-falling particles
with negative energy, one could assume the dimensionality of the whole Hilbert
space remains unchanged in the radiation process. The creation of the pair
outside the horizon and the annihilation of the particle inside the horizon
are treated only as intermediate processes to accommodate for the change of an
initial black hole into a reduced black hole plus radiation. Other discussions
prefer to presume an initial Hilbert space as consisting of the black hole and
its exterior \cite{tb11,sbg12}, which might be easier to arrive at a unitary
process with a semiclassical thermal black hole radiation.

\section{Correlation and Entropy Conservation}

In order to resolve the paradox with the non-thermal spectrum (\ref{nt}), the
first question we have to ask is whether there exist correlations among the
non-thermal radiations. This same question was initially addressed by Parikh
\cite{mkp04} and then by Arzano \textit{et al }\cite{amv05} with a negative
answer, i.e., both found existence of no correlations. In our recent work, we
find, however, that their answers are incorrect and correlations indeed exist
among Hawking radiations from a non-thermal spectrum.

So what is the correct answer? For that, we go to the statistical theory.
Given two statistical events, like two emissions of Hawking radiation, with
their joint probability denoted by $p(A,B)$, the probabilities for each single
emissions are given by $p(A)=\int p(A,B)dB$ and $p(B)=\int p(A,B)dA$. We then
proceed with a simple check to see whether $p(A,B)=p(A)\cdot p(B)$ holds true
or not. If the equality sign holds true, no correlation exists between event
$A$ and event $B$. The two events are independent. This is indeed the case
when the emission spectrum is thermal. For the non-thermal spectrum of Eq.
(\ref{nt}), the two emissions are dependent, thus correlated. Alternatively,
we can check if the conditional probability $p(B|A)=p(A,B)/p(A)$ of event $B$
to occur given that event $A$ has occurred is equal to the probability $p(B)$
\cite{gs92}.

As we have previously shown, when considering two subsequent emissions with
energies $\mathit{E}_{1}$ and $\mathit{E}_{2}$, the tunneling probability for
a particle of energy $\mathit{E}_{2}$ has to be treated carefully. Since we
cannot assume in advance whether there exists correlation between the two
emissions or not, we have to get the two probabilities by taking the integral
of their joint probability $\Gamma\left(  \mathit{E}_{1},\mathit{E}%
_{2}\right)  $, i.e. $\Gamma\left(  \mathit{E}_{1}\right)  =\int\Gamma\left(
\mathit{E}_{1},\mathit{E}_{2}\right)  d\mathit{E}_{2}$ and $\Gamma\left(
\mathit{E}_{2}\right)  =\int\Gamma\left(  \mathit{E}_{1},\mathit{E}%
_{2}\right)  d\mathit{E}_{1}$. The joint probability refers to the two
emissions with energies $\mathit{E}_{1}$ and $\mathit{E}_{2}$ occuring
simultaneously, i.e., we have
\[
\Gamma\left(  \mathit{E}_{1},\mathit{E}_{2}\right)  =\Gamma\left(
\mathit{E}_{1}+\mathit{E}_{2}\right)  ,
\]
which holds true for the non-thermal spectrum of Eq. (\ref{nt}) as energy
conservation is enforced for its derivation within the treatment of Hawking
radiation as tunneling. Then we get the two independent tunneling
probabilities respectively as%
\begin{align*}
\Gamma\left(  \mathit{E}_{1}\right)   &  =\exp\left[  -8\pi\mathit{E}%
_{1}\left(  M-\frac{\mathit{E}_{1}}{2}\right)  \right]  ,\\
\Gamma\left(  \mathit{E}_{2}\right)   &  =\exp\left[  -8\pi\mathit{E}%
_{2}\left(  M-\frac{\mathit{E}_{2}}{2}\right)  \right]  .
\end{align*}
Thus on one hand, we can affirm the existence of correlation by finding the
conditional probability $\Gamma\left(  \mathit{E}_{2}|\mathit{E}_{1}\right)
=\frac{\Gamma\left(  \mathit{E}_{1},\mathit{E}_{2}\right)  }{\Gamma\left(
\mathit{E}_{1}\right)  }\neq\Gamma\left(  \mathit{E}_{2}\right)  $, and on the
other hand, due to $\Gamma\left(  \mathit{E}_{1},\mathit{E}_{2}\right)
\neq\Gamma\left(  \mathit{E}_{1}\right)  \cdot\Gamma\left(  \mathit{E}%
_{2}\right)  $, we can define a quantity to measure the correlation as in Ref.
\cite{amv05},%
\begin{equation}
C\left(  \mathit{E}_{1}+\mathit{E}_{2};\mathit{E}_{1},\mathit{E}_{2}\right)
=\ln\Gamma\left(  \mathit{E}_{1},\mathit{E}_{2}\right)  -\ln\left[
\Gamma\left(  \mathit{E}_{1}\right)  \cdot\Gamma\left(  \mathit{E}_{2}\right)
\right]  . \label{cq}%
\end{equation}
A simply calculation shows the correlation between two emissions is
$8\pi\mathit{E}_{1}\mathit{E}_{2}$; we can proceed to calculate the
correlation between these two emissions with a third one of energy
$\mathit{E}_{3}$, and the resulting correlation is found to be $8\pi\left(
\mathit{E}_{1}+\mathit{E}_{2}\right)  \mathit{E}_{3}$; the total correlation
among the three emissions $\mathit{E}_{1}$, $\mathit{E}_{2}$ and
$\mathit{E}_{3}$ thus becomes $8\pi\mathit{E}_{1}\mathit{E}_{2}+8\pi\left(
\mathit{E}_{1}+\mathit{E}_{2}\right)  \mathit{E}_{3}$, which does not depend
on the orders of the individual emissions. So for the non-thermal spectrum,
the total correlation in a queue of Hawking radiations is summed up to%
\begin{equation}
C\left(  M;\mathit{E}_{1},\mathit{E}_{2}\cdots\mathit{E}_{n}\right)
=\sum_{n\geqslant2}8\pi\left(  \mathit{E}_{1}+\mathit{E}_{2}+\cdots
+\mathit{E}_{n-1}\right)  \mathit{E}_{n}=\sum_{i\geqslant j}8\pi\mathit{E}%
_{i}\mathit{E}_{j} \label{tcq}%
\end{equation}
where indices $i$ and $j$ take all possible values labeling individual
emissions. The analogous correlation are found to vanish for a thermal spectrum.

Another proper description for the correlation appeals to the concept of the
mutual information which is defined for a composite quantum system composed of
subsystems $A$ and $B$ as%
\[
S(A:B)\equiv S(A)+S(B)-S(A,B)=S(A)-S(A|B),
\]
where $S(A|B)$ is nothing but the conditional entropy \cite{nc00}. This can be
used to measure the total amount of correlations between any bi-partite
system. To use the mutual information, we have to introduce the entropy for
the discussed events here, i.e. the tunneling emission of particles. From our
earlier analysis, the entropy taken away by the tunneling particle with energy
$\mathit{E}_{i}$ after the black hole has emitted particles with a total
energy $\mathit{E}_{f}=\overset{i-1}{\underset{j=1}{\sum}}\mathit{E}_{j}$ is
given by
\begin{equation}
S\left(  \mathit{E}_{i}|\mathit{E}_{1},\mathit{E}_{2},\ldots,\mathit{E}%
_{i-1}\right)  =-\ln\Gamma(\mathit{E}_{i}|\mathit{E}_{1},\mathit{E}_{2}%
,\ldots\mathit{E}_{i-1}). \label{ce1}%
\end{equation}
Again $S\left(  \mathit{E}_{i}|\mathit{E}_{1},\mathit{E}_{2},\ldots
,\mathit{E}_{i-1}\right)  $ denotes the conditional entropy that measures the
entropy of emission $\mathit{E}_{i}$ given that the values of all the emitted
particles with energies $\mathit{E}_{1},\mathit{E}_{2}$, $\ldots$, and
$\mathit{E}_{i-1}$ are known. Quantitatively, it is equal to the decrease of
the entropy of a black hole with mass $M-\mathit{E}_{f}$ upon the emission of
a particle with energy $\mathit{E}_{i}$. Such a result is also consistent with
the thermodynamic second law for a black hole \cite{swh760}: the emitted
particles must carry entropies in order to balance the total entropy of the
black hole and the radiation. When mutual information is applied to the
emissions of two particles with energies $\mathit{E}_{1}$ and $\mathit{E}_{2}%
$, we have%
\begin{equation}
S(\mathit{E}_{2}:\mathit{E}_{1})\equiv S(\mathit{E}_{2})-S(\mathit{E}%
_{2}|\mathit{E}_{1})=-\ln\Gamma(\mathit{E}_{2})+\ln\Gamma(\mathit{E}%
_{2}|\mathit{E}_{1}). \label{tmi}%
\end{equation}
Thus it is easy to find $S(\mathit{E}_{2}:\mathit{E}_{1})=8\pi\mathit{E}%
_{1}\mathit{E}_{2}$, which shows the correlation of Eq. (\ref{cq}) is exactly
equal to the mutual information between the two sequential emissions.

The conditional entropy (\ref{ce1}) can be used to calculate the total entropy
for the radiations after a black hole evaporates, which is
\begin{equation}
S(\mathit{E}_{1},\mathit{E}_{2},\cdots,\mathit{E}_{n})=\sum\limits_{i=1}%
^{n}S(\mathit{E}_{i}|\mathit{E}_{1},\mathit{E}_{2},\cdots,\mathit{E}_{i-1}),
\label{te}%
\end{equation}
where $M=\sum_{i=1}^{n}\mathit{E}_{i}$ equals to the initial black hole mass
due to energy conservation and $S(\mathit{E}_{1},\mathit{E}_{2},...,\mathit{E}%
_{n})$ denotes the joint entropy for all emissions. By a detailed calculation
from Eq. (\ref{te}), we previously show that the total entropy $S(\mathit{E}%
_{1},\mathit{E}_{2},...,\mathit{E}_{n})=4\pi M^{2}$ exactly equals to the
black hole's Bekenstein-Hawking entropy. This shows that the entropy of a
black hole is indeed taken out by Hawking radiations, and the total entropy of
all emitted radiations and the black hole is conserved during the black hole
radiation process, which shows the consistency of Hawking radiation with the
unitarity of quantum mechanics. Alternatively, this result can be understood
by counting the number of ways a black hole can evaporate as in our earlier
analysis \cite{zcyz09} and in a later study by Israel and Yun \cite{iy10};
i.e. the probability for evaporation of a black hole can be expressed as%
\[
\Gamma\left(  M\rightarrow M-\mathit{E}_{1}\rightarrow M-\mathit{E}%
_{1}-\mathit{E}_{2}\rightarrow\cdots\rightarrow0\right)  =e^{-S_{\mathrm{BH}}%
}.
\]
The number of ways for evaporation is then given by $N=$ $e^{S_{\mathrm{BH}}}$
because $N\Gamma=1$, which provides an easy interpretation of the
Bekenstein-Hawking entropy $S_{\mathrm{BH}}$ in terms of the number of modes
for evaporation. In particular, at each step, the tunneling probability
manifests itself as a quantum transition probability, i.e. $\Gamma\sim
\frac{e^{S_{\mathrm{fianl}}}}{e^{S_{\mathrm{initial}}}}$, up to leading order.

\section{Entropy}

In the previous section, we briefly reviewed our earlier results that there
exist correlations among Hawking radiations treated as quantum tunneling. In
the radiation process the total entropy for a black hole and its radiations is
unchanged, which thus establishes its consistency with the unitarity of
quantum mechanics. Although an interpretation of the Bekenstein-Hawking
entropy is provided in terms of the number of modes for a black hole's
evaporation, the specific meaning of the entropy referred to Hawking radiation
remains unclear from the discussions in the above. This puzzling point was
coincidentally treated by us in another earlier study \cite{zczy113}, where we
interpret the above discussed entropy in terms of the uncertainty about the
information of the precollapsed configurations of a black hole's forming
matter, the self-collapsed configurations and the inter-collapsed
configurations. Our interpretation can be applied to several relevant
circumstances, including the formation of a black hole, the black hole
coalescence, and a common matter dropped into a black hole.

Our interpretation can be presented using an explicit expression, i.e. the
entropy carried away by an emitted particle with energy $\mathit{E}$,%
\begin{equation}
S\left(  \mathit{E}\right)  =8\pi\mathit{E}\left(  M-\mathit{E}\right)
+\left(  4\pi\mathit{E}^{2}-S_{0}\right)  +S_{0}, \label{pte}%
\end{equation}
where the entropy is partitioned into three parts with $S_{0}$ referring to
the precollapsed configuration which reveals the information about the matter
that will collapse into a black hole, $\left(  4\pi\mathit{E}^{2}%
-S_{0}\right)  $ about self-collapsed configuration which reveals the
inaccessible information about how to collapse, and the correlation or the
partial information $8\pi\mathit{E}\left(  M-\mathit{E}\right)  $ about
inter-collapsed configuration which reveals that the inaccessible information
about the interaction among different collapsed holes. Maybe it is better to
state the entropy in another way, i.e. it means that the tunneling particle
carries away all information about its precollapsed configuration, the
self-collapsed configuration, and the inter-collapsed configuration.

The above partition spells out what we interpret the meaning of the
Bekenstein-Hawking entropy from the radiations. Initially this insight was
gained from observing a hypothetical collision of two Schwarzschild black
holes with respective masses $m_{1}$ and $m_{2}$. Their respective entropies
are $4\pi m_{1}^{2}$ and $4\pi m_{2}^{2}$. If the two black holes collide and
form a new black hole with mass $m=m_{1}+m_{2}$ as required from energy
conservation, the entropy for the new black hole becomes $S_{m}=4\pi\left(
m_{1}+m_{2}\right)  ^{2}=4\pi m_{1}^{2}+4\pi m_{2}^{2}+8\pi m_{1}m_{2}$ which
is not equal to the sum of the entropies of the two initial black holes. This
means correlations generated by gravitational interaction arise and are hidden
inside the newly formed black hole and the exterior observer cannot obtain any
information about them. On the other hand, when the two black holes collide
and coalesce into one, gravitational waves will usually be emitted in the
process. Is it possible that gravitational radiations carry away the amount of
information related to the term $8\pi m_{1}m_{2}$ as constrained by entropy
conservation? Our earlier analysis \cite{zczy113} show that this is
impossible. Actually Hawking radiation is probably the best messenger to carry
away information locked inside a black hole according to our entropy partition
(\ref{pte}). In particular, we have shown earlier that the information carried
away does not include any redundant contents by considering a queue of
emissions ordered according to $\mathit{E}_{1}$, $\mathit{E}_{2}$, $\cdots$,
$\mathit{E}_{n-1}$, and $\mathit{E}_{n}$. The entropy for the first emission
with an energy $\mathit{E}_{1}$ is%
\begin{equation}
S\left(  \mathit{E}_{1}\right)  =8\pi\mathit{E}_{1}\left(  M-\mathit{E}%
_{1}\right)  +\left(  4\pi\mathit{E}_{1}^{2}-S_{01}\right)  +S_{01}%
\end{equation}
where the term $8\pi\mathit{E}_{1}\left(  M-\mathit{E}_{1}\right)  $ includes
all correlations between the particle with energy $\mathit{E}_{1}$ and all
other particles with energies $\mathit{E}_{2}$, $\cdots$, $\mathit{E}_{n-1}$,
and $\mathit{E}_{n}$. The entropy for the second emission with an energy
$\mathit{E}_{2}$ given the first emission with an energy $\mathit{E}_{1}$ is%
\begin{equation}
S\left(  \mathit{E}_{2}|\mathit{E}_{1}\right)  =8\pi\mathit{E}_{2}\left(
M-\mathit{E}_{1}-\mathit{E}_{2}\right)  +\left(  4\pi\mathit{E}_{2}^{2}%
-S_{02}\right)  +S_{02}%
\end{equation}
where we see that information about the interaction between the particles with
energies $\mathit{E}_{1}$ and $\mathit{E}_{2}$ is taken out by the first
emission. For the second emission, we therefore must subtract the quantity
$8\pi\mathit{E}_{1}\mathit{E}_{2}$ from the information it will take out.
Analogously, for the entropy of the third emission with energy $\mathit{E}%
_{3}$, we find%
\begin{equation}
S\left(  \mathit{E}_{3}|\mathit{E}_{1},\mathit{E}_{2}\right)  =8\pi
\mathit{E}_{3}\left(  M-\mathit{E}_{1}-\mathit{E}_{2}-\mathit{E}_{3}\right)
+\left(  4\pi\mathit{E}_{3}^{2}-S_{03}\right)  +S_{03},
\end{equation}
again the correlations between the third emission $\mathit{E}_{3}$ with the
first two of energies $\mathit{E}_{1}$ and $\mathit{E}_{2}$ need to be
subtracted. Summing up the three entropies, we find there exists no redundant
information or entropy
\begin{align}
&  S\left(  \mathit{E}_{1}\right)  +S\left(  \mathit{E}_{2}|\mathit{E}%
_{1}\right)  +S\left(  \mathit{E}_{3}|\mathit{E}_{1},\mathit{E}_{2}\right)
\nonumber\\
&  =4\pi\left(  M-\mathit{E}_{1}-\mathit{E}_{2}-\mathit{E}_{3}\right)
^{2}-4\pi M^{2}\nonumber\\
&  =\Delta S_{\mathrm{BH}}.
\end{align}
A step by step follow up calculation show that according to our suggested
partition (\ref{pte}) not only that Hawking radiations carry with themselves
all information with no loss or redundance. Thus it provides a self-consistent
interpretation for the entropy of a black hole. It implies to an exterior
observer, there exists an uncertainty or information about the back hole's
precollapsed configuration, its self-collapsed configuration, and
inter-collapsed configuration.

For the final radiation, there still remains some indefiniteness in the
process of Hawking radiation as tunneling, since its entropy is given by
\begin{equation}
S\left(  \mathit{E}_{n}|\mathit{E}_{1},\mathit{E}_{2}\cdots,\mathit{E}%
_{n-1}\right)  =4\pi\mathit{E}_{n}^{2}%
\end{equation}
which is precisely the same as for a black hole with mass or energy
$\mathit{E}_{n}$. This shows the final emission is really equivalent to no
emission or emit itself. In other words, the Hawking radiation as tunneling
cannot give any better description for the final emission than before, which
can be compared to the views of considering the final black hole as a
fundamental particle \cite{gct85,ykh09} or a stable remnant \cite{amv05,lx07}.

A second puzzling question contrasts how the entropy for an ordinary matter,
which could essentially take any value, changes into a fixed value
$4\pi\mathit{E}^{2}$ after transformed into a black hole of an equal energy
$\mathit{E}$? A clear decisive answer remains lacking here. However, we can
shed some light on its answer using a conjecture by some physicists
\cite{jdb81,ls95,rb99,fmw00}, which claims black holes have the maximum
possible entropy of any object of equal size and as such makes them the likely
end points of all entropy-increasing processes. For a clearer interpretation
about the black hole entropy, we need a better description for the state of
the black hole's interior.

\section{Information Transfer}

In the above sections, we have presented formally a mechanism about how to
preserve unitarity in the process of Hawking radiation as tunneling, with the
discovery of correlation among radiations from a non-thermal spectrum to
balance the otherwise ever increasing entropies of Hawking radiations.
Although we have pointed out implicitly that information can be taken out by
Hawking radiation, the explicit mechanism for transfer (e.g. how information
is encoded and decoded in the process of black hole collapse and radiation) is
still unclear. In this section, we will give two examples to show that
information hidden in a black hole can indeed be carried away by correlations
among Hawking radiations, irrespective what the specific transfer mechanism is.

Our first example concerns Hawking radiation as tunneling through a quantum
horizon \cite{zczy11,cs09}. The tunneling probability was already given
\cite{amv05} for a general spherically symmetric system in the ADM form
\cite{kw95} by referencing to the first law of black hole thermodynamics
$dM=\frac{\kappa}{2\pi}dS$,
\begin{equation}
\Gamma\sim\left(  1-\frac{\mathit{E}}{M}\right)  ^{2\alpha}\exp\left[
-8\pi\mathit{E}\left(  M-\frac{\mathit{E}}{2}\right)  \right]  =\exp\left(
\Delta S\right)  , \label{qtpo}%
\end{equation}
where $S=\frac{A}{4}+\alpha\ln A$ is the entropy derived by directly counting
the number of micro-states with string theory and loop quantum gravity
\cite{amv05}. The coefficient $\alpha$ is negative in loop quantum gravity
\cite{gm05}. Its sign remains uncertain in string theory, depending on the
number of field species in the low energy approximation \cite{sns98}. For
$\alpha>0$, we find $\Gamma\rightarrow0$ when $\mathit{E}\rightarrow M$, but
$S\left(  M-\mathit{E}\right)  \rightarrow\infty$. This causes difficulty in
explaining the origin of an exponentially growing entropy when the black hole
vanishes. However, qualitatively, this actually can be understood within the
picture of Hawking radiation from a black hole. In the limit of $\Gamma(M)$
$=0$, the tunneling energy approaches the mass of the black hole, and the
tunneling becomes slower and slower while the time to exhaust a black-hole
approaches infinite. This infinity can also be obtained from other methods by
using the Stefan-Boltzmann law as in Ref. \cite{fp07}. For $\alpha<0$, it is
known \cite{lx07} that when the mass of a black hole approaches the critical
mass $M_{c}$, no particles will be emitted. Using our previous analysis as
presented in Eq. (\ref{te}), one then obtain $S(M)-S(M_{c})=\sum
\limits_{i}S(\mathit{E}_{i}|\mathit{E}_{1},\mathit{E}_{2},\cdots
,\mathit{E}_{i-1})$ or $S(M)=\sum\limits_{i}S(\mathit{E}_{i}|\mathit{E}%
_{1},\mathit{E}_{2},\cdots,\mathit{E}_{i-1})+S(M_{c})$. In Ref. \cite{lx07},
the mass $M_{c}$ is called the \textquotedblleft zero point
energy\textquotedblright\ of a black hole that is similar to a black hole
remnant because it does not depend on the initial black hole mass. We showed
that even with such a remnant, the total entropy remains conserved when
information carried away by correlations are correctly included. Thus the
unitarity remains true when the classical horizon is replaced by a quantum one
for Hawking radiation.

More significantly, we want to affirm whether information about quantum
gravity corrections or about $\alpha$ is taken out by Hawking radiations. This
can be confirmed through the correlation%
\begin{equation}
C\left(  \mathit{E}_{1}+\mathit{E}_{2};\mathit{E}_{1},\mathit{E}_{2}\right)
=8\pi\mathit{E}_{1}\mathit{E}_{2}+2\alpha\ln\frac{M(M-\mathit{E}%
_{1}-\mathit{E}_{2})}{(M-\mathit{E}_{1})(M-\mathit{E}_{2})}\neq0. \label{qco}%
\end{equation}
which shows that the information about quantum gravity effect is actually
carried away if quantum gravity effects indeed exist in the interior of a
black hole.

The second example concerns noncommutative space \cite{bms08,zczy112}. In
order to include the noncommutative space effect in gravity, one has to change
the mass of a gravitating object. The usual definition of mass density in
commutative space is expressed in terms of Dirac delta function, but in
noncommutative space this form breaks down due to position-position
uncertainty relations. It was shown \cite{ss031,ss032} that noncommutativity
eliminates point-like structures in favor of smeared objects in flat
spacetime. The effect of smearing is implemented by redefining the mass
density by a Gaussian distribution of a minimal width $\sqrt{\theta}$ instead
of the Dirac delta function. Here $\theta$ is the noncommutative parameter
which is considered to be a small (Planck length) positive number and comes
from the noncommutator of $\left[  x^{\mu},x^{\upsilon}\right]  =i\theta
^{\mu\nu}$ with $\theta^{\mu\nu}=\theta\,\mathrm{diag}\left[  \epsilon
_{1},\cdots,\epsilon_{D/2}\right]  $. The constancy of $\theta$ is related to
a consistent treatment of Lorentz invariance and unitarity \cite{ss04}.

Under the assumption of spatial noncommutativity, the background Painlev\'{e}
coordinate becomes \cite{zczy112,bms08,nss06}%
\begin{equation}
ds^{2}=-\left(  1-\frac{4M}{r\sqrt{\pi}}\gamma\right)  dt^{2}+2\left(
1-\frac{4M}{r\sqrt{\pi}}\gamma\right)  \sqrt{\frac{\frac{4M}{r\sqrt{\pi}%
}\gamma}{\left(  1-\frac{4M}{r\sqrt{\pi}}\gamma\right)  ^{2}}}\,dtdr+dr^{2}%
+r^{2}d\Omega^{2}, \label{nspc}%
\end{equation}
where $\gamma\equiv\gamma\left(  \frac{3}{2},\frac{r^{2}}{4\theta}\right)  $
and the spacetime described by (\ref{nspc}) is still stationary. Then a
detailed calculation gives the tunneling rate,%
\begin{equation}
\Gamma\sim\exp\left[  -8\pi\mathit{E}\left(  M-\frac{\mathit{E}}{2}\right)
+16\sqrt{\frac{\pi}{\theta}}M^{3}e^{\frac{M^{2}}{\theta}}-16\sqrt{\frac{\pi
}{\theta}}\left(  M-\mathit{E}\right)  ^{3}e^{\frac{\left(  M-\mathit{E}%
\right)  ^{2}}{\theta}}\right]  , \label{nst}%
\end{equation}
where the result is obtained up to the leading order $\frac{1}{\sqrt{\theta}%
}e^{-M^{2}/\theta}$. With the same procedures presented in the third section,
one can prove the total entropy is still conserved in the radiation process
with the background metric (\ref{nspc}). In particular, we find the
statistical correlation between two emissions
\begin{equation}
C\left(  \mathit{E}_{1}+\mathit{E}_{2};\mathit{E}_{1},\mathit{E}_{2}\right)
=C\left(  \mathit{E}_{1},\mathit{E}_{2},\theta\right)  \neq0,
\end{equation}
which shows that information about the spatial noncommutativity labeled by the
parameter $\theta$ will be carried out if space is indeed noncommutative.

A recent paper \cite{gs13} studied information transfer mechanism by
attempting to match a black hole's evolution unitarily. Due to the complexity
of black hole physics, e.g. it remains unknown what dynamics is really
responsible for the description of a black hole's interior and most physicists
even believe that physics near the event horizon (about the Planck length away
from the horizon) cannot be properly described with quantum field theory, no
workable information transfer mechanism has been found capable of fitting in
with the black hole evolution to the presumed being unitarity, especially
after the paradox of \textquotedblleft firewall\textquotedblright%
\ \cite{amps13} was put forward. Nevertheless, it is found that our
description for Hawking radiation is the same as the mechanisms called
\textquotedblleft subsystem transfer\textquotedblright\ \cite{gs13}, since
within each step the sum of the remaining black hole's entropy and the entropy
carried away by the outgoing particles is exactly equal to the initial entropy
of the black hole. This not only satisfies the condition required by
information transfer, i.e. the entropy for a black hole decreases, it also
saturates the subadditivity inequality required by the subsystem transfer,
although we didn't assume in advance the existence of external Hilbert space
as was done in the Ref. \cite{gs13}. Maybe Page's work \cite{dnp93} with the
Hilbert space consisting of a black hole and its radiation subsystem is closer
to our description. But in his model the final entropy of radiation subsystem
is found to decrease to zero.

\section{Conclusion}

In this brief review, we have presented our earlier studies and conclusions
regarding the black hole information loss paradox. After our discovery that
the non-thermal spectrum of Parikh and Wilczek allows for the Hawking
radiation emissions to carry off all information of a black hole, a natural
question to ask is whether Hawking radiation is indeed non-thermal or not?
Although the derivation of the non-thermal spectrum is based on solid physics,
it remains to be confirmed experimentally or in observations. A recent
analysis by us \cite{zcy13} show that the non-thermal spectrum can indeed be
distinguished from the thermal spectrum by counting the energy covariances of
Hawking radiations. The energy covariances actually measure the correlation
among Hawking radiations. With the relatively low energy scale for quantum
gravity and the large dimensions, the production of micro black holes and the
observation of Hawking radiations has already been studied
\cite{bf99,cms11,mr08,gt02,dl01,ehm00}. If Hawking radiations from a micro
black hole were observed in an LHC experiment, our results show that it can
definitely determine whether the emission spectrum is indeed non-thermal
\cite{zcy13} or not. Thus it provides an avenue towards experimentally testing
the long-standing \textquotedblleft information loss paradox\textquotedblright%
. On the other hand, as shown in the previous section, the correlation we
discuss is also the carrier of information hidden in the interior of a black
hole, so if the correlation is found experimentally, more information about
some fundamental theories such as quantum gravity might be revealed simultaneously.

We conclude that our studies show the existence of correlations among Hawking
radiations and the total entropy is conserved in the whole radiation process,
which is consistent with unitarity of quantum mechanics. Our series of studies
resolve the paradox of information loss at least formally, although the
framework of Hawking radiation as quantum tunneling constrains our ability to
explore the internal dynamics of a black hole. We have also made some remarks
and comments to several recent researches on the subject, and found no
inconsistencies with our results. Although our recent work widely supports the
claim that Hawking radiation process is not in conflict with quantum mechanics
at the semiclassical level, a more refined description for the radiation
process remains to be constructed. Further study along this direction could
possibly confront the emergence of new physics. Irrespective of that, however,
as we declared in a recent essay \cite{zczy132}, which won the first prize in
the 2013 Essay Competition of the Gravity Research Foundation, information
conservation is fundamental for any isolated system, even for a black hole
with its radiations being part of the system.

\section{Acknowledgement}

Financial support from National Natural Science Foundation of China under
Grant Nos. 11104324, 11374330, 11074283, 11227803, 11374176, and 91121005.

\end{document}